\documentclass[12pt,preprint]{aastex}

\usepackage{url}

\shorttitle{Sky brightness and transparency at Dome A}
\shortauthors{Hu Zou, Xu Zhou et al.}

\begin{document}

\title{The sky brightness and transparency in $i$-band at Dome A, Antarctica}
\author{Hu Zou\altaffilmark{1,2,3}, Xu Zhou\altaffilmark{1,3},
  Zhaoji Jiang\altaffilmark{1,3},   M.~C.~B.~Ashley\altaffilmark{4},
  Xiangqun Cui\altaffilmark{3,5},  Longlong Feng\altaffilmark{3,6},
  Xuefei Gong\altaffilmark{3,5}, Jingyao Hu\altaffilmark{1,3},
  C.~A.~Kulesa\altaffilmark{7},  J.~S.~Lawrence\altaffilmark{8,9},
  Genrong Liu\altaffilmark{5}, D.~M.~Luong-Van\altaffilmark{4},
  Jun Ma\altaffilmark{1}, A.~M.~Moore\altaffilmark{10},
  C. R. Pennypacker\altaffilmark{11}, Weijia Qin\altaffilmark{12}, Zhaohui Shang\altaffilmark{13},
  J.~W.~V.~Storey\altaffilmark{4}, Bo Sun,\altaffilmark{12},
  T.~Travouillon\altaffilmark{11}, C.~K.~Walker\altaffilmark{7},
  Jiali Wang\altaffilmark{1,3}, Lifan Wang\altaffilmark{3,6},
  Jianghua Wu\altaffilmark{1}, Zhenyu Wu\altaffilmark{1},
  Lirong Xia\altaffilmark{5}, Jun Yan\altaffilmark{1,3},
  Ji Yang\altaffilmark{6}, Huigen Yang\altaffilmark{12},
  Yongqiang Yao\altaffilmark{1}, Xiangyan Yuan\altaffilmark{3,5},
  D.~G.~York\altaffilmark{14},  Zhanhai Zhang\altaffilmark{12},
  Zhenxi Zhu\altaffilmark{3,6}}

\altaffiltext{1}{National Astronomical Observatories, Chinese Academy of Sciences, Beijing 100012, China;\\ zhouxu@bao.ac.cn}
\altaffiltext{2}{Graduate University of Chinese Academy of Sciences, Beijing 100049, China}
\altaffiltext{3}{Chinese Center for Antarctic Astronomy, Purple Mountain Observatory, Chinese Academy of Sciences, Nanjing 210008, China}
\altaffiltext{4}{School of Physics, University of New South Wales, NSW 2052, Australia}
\altaffiltext{5}{Nanjing Institute of Astronomical Optics and Technology, Nanjing 210042, China}
\altaffiltext{6}{Purple Mountain Observatory, Chinese Academy of Sciences, Nanjing 210008, China}
\altaffiltext{7}{Steward Observatory, University of Arizona, Tucson, AZ 85721, USA}
\altaffiltext{8}{Department of Physics, Macquarie University, NSW 2109, Australia}
\altaffiltext{9}{Anglo-Australian Observatory, NSW 1710, Australia}
\altaffiltext{10}{Department of Astronomy, California Institute of Technology, Pasadena, CA 91125, USA}
\altaffiltext{11}{Lawrence Berkeley National Laboratory, Berkeley, CA 94720, USA}
\altaffiltext{12}{Polar Research Institute of China, Pudong, Shanghai 200136, China}
\altaffiltext{13}{Tianjin Normal University, Tianjin 300074, China}
\altaffiltext{14}{Department of Astronomy and Astrophysics and Enrico Fermi Institute, University of Chicago, Chicago, IL 60637, USA}

\begin{abstract}
The $i$-band observing conditions at Dome A on the Antarctic plateau
have been investigated using data acquired during 2008 with the
Chinese Small Telescope ARray. The sky brightness,
variations in atmospheric transparency, cloud cover, and the presence of aurorae are
obtained from these images. The median sky brightness of moonless clear
nights is 20.5 mag arcsec$^{-2}$ in the SDSS $i$ band
at the South Celestial Pole (which includes a contribution of about 0.06 mag
from diffuse Galactic light). The median over all Moon phases in the Antarctic
winter is about 19.8 mag arcsec$^{-2}$. There were no thick clouds in 2008.
We model contributions of the Sun and the Moon to the sky background to obtain the relationship
between the sky brightness and transparency. Aurorae are identified by
comparing the observed sky brightness to the sky brightness expected
from this model. About 2\% of the images are affected by relatively strong aurorae.
\end{abstract}


\section{INTRODUCTION}
In selecting observatory sites for ground-based optical/IR astronomy, some of the most
important considerations are the night-sky brightness,
transparency, seeing, number of clear nights, humidity, and
photometric stability. The Antarctic plateau offers some
attractive advantages for ground-based astronomical observations. Site
testing over the past decade has revealed that Antarctica, relative to
temperate latitude observatories, has lower
infrared sky brightness, better free-atmosphere seeing, greater
transparency, a lower turbulent boundary layer, and much lower water
vapor content (see, e.g., reviews by \citet{ari05a},\citet{bur05}, and \citet{sto07}).
About half the area of
the Antarctic continent is at an elevation of over 3000 m above the
sea level, and the year-round average temperature on the plateau
is about $-50^\circ \mathrm{C}$, dropping to below $-80^\circ
\mathrm{C}$ on occasion. The average of the water vapor column over
the plateau has been predicted to be less than 250 $\mu$m \citep{cha01}
and has been measured
at Dome A to be below 100 $\mu$m for 25\% of the time \citep{yan10}.
Wind speeds at the summits of
the plateau are very low close to the surface and the median
thickness of the turbulent boundary layer has been measured at Dome A as 13.9~m
\citep{bon10}. These exceptionally favorable geographical and
meteorological conditions result in substantial advantages for millimeter,
sub-millimeter, infrared, and optical astronomy.

Several large astronomical facilities are already operating at
some sites on the Antarctic plateau, e.g., the South Pole
Telescope \citep{ruh04} at the US Amundsen-Scott station.
Proposed facilities include a 2.5 m optical telescope, PILOT (now called PLT)
at Concordia Station, Dome C \citep{sto08,law09b}, and a 2 m class
infrared telescope planned for deployment to Dome Fuji by Japan
\citep{tak08}.

At Dome C (3250 m elevation), the average seeing is 0.\arcsec27 above
a $\sim30$ m ground layer, and the average surface wind speed is 2.9 m s$^{-1}$
\citep{law04,ari05a,ari05b}. Dome A is the
highest point of the continent, with an elevation of 4093~m. It might be
reasonably predicted that Dome A could be as good as or even a better astronomical site than Dome
C, with better seeing, higher transparency, and thinner surface layer.
\citet{sau09} compared the sites Dome A, B, C, F and Ridge
A and B in their cloud cover, free-atmosphere seeing, precipitable water
vapor, temperature, and auroral emission, and concluded that,
overall, Dome A might be the best of the existing bases for
astronomical observations.

Following the success of the first Chinese expedition team to Dome A
in 2005 January, the Chinese Center for Antarctic Astronomy (CCAA)
began planning for the deployment of a telescope array called Chinese
Small Telescope ARray (CSTAR; \citet{yua08}) described in the following section.

In this paper, we use CSTAR images obtained during 2008 to assess the
night sky brightness at Dome A. Both the variation in atmospheric
transparency and cloud cover are also derived. We then analyze the
contribution to the sky brightness from the Sun and the Moon and
investigate the correlation between sky brightness and atmospheric transparency.
Finally, the contribution from aurorae is assessed.

\section{INSTRUMENT AND OBSERVATIONS}

The CSTAR telescope array was built by the Nanjing Institute of
Astronomical Optics \& Technology (NIAOT) and consists of four
co-aligned Schmidt telescopes on a fixed mounting pointing toward the
South Celestial Pole. Each telescope has a different filter: $g$, $r$,
$i$, and open. The detectors are Andor DV435
1k$\times$1k CCDs with 13 $\mu$m pixels, giving a plate scale
of 15\arcsec\ pixel$^{-1}$. The entrance pupil diameter is 145mm (effective
aperture: 100mm) and the field of view is about 20 deg$^2$.  The
scientific goals for CSTAR were primarily the detection of variable
stars and measurements of sky brightness, the stability of the
atmospheric transparency, and cloud cover.

The CSTAR telescopes, cameras, and computer system were tested under
conditions of very low temperature, low air pressure, and
long-duration continuous operation. The observational testing of the array
was done at the Xinglong Station of National Astronomical Observatories
of China (NAOC; \citet{zho10a}).

In 2008 January, CSTAR was transported to Dome A and
successfully installed, with the intention of making
observations of the South Celestial Pole region from 2008 April to August.
Power, heat, and Internet connectivity for CSTAR were provided by
PLATO (the PLATeau Observatory), developed by the University
of New South Wales \citep{yan09,law09a}.

After deployment to Dome A, only the $i$-band
telescope (CSTAR\#1) operated well, observing throughout the winter
with just a few short interruptions; the other three telescopes had either small problems with their
optical alignment or their control computers.
However, all four telescopes operated sufficiently well to allow us to derive some
information on sky conditions in the other filter bands, and this
analysis will be published in a subsequent paper.
Over 310,000 images were obtained, with integration times of 20 s or 30 s,
between 2008 March 4 and 2008 August 8, giving a total exposure time
on the sky of about 1728 hr \citep{zho10b}. While
CSTAR was observing, roughly one-third of the catalogs and one
original image were transferred via Iridium communication satellite
each day. All the catalogs and images were brought back by
the 25th Chinese Research Expedition Team, in 2009
April. The catalogs contain the ambient temperature, observing
date and time, exposure time and background brightness, plus the
instrumental magnitudes in three different apertures for all
detected point sources calculated by an automatic aperture
photometry pipeline. For each image about 10,000 sources down to $i\sim$16
mag were detected. Several variable stars were identified and the
complete observational catalog has now been released
\citep{zho10b}. As the catalog has continuous
observations over several months, the data are ideal for the
detection and monitoring of variable stars (J. H. Wu et al. 2010, in preparation).

Observations began on 2008 March 4 and ended on 2008 August 8, a few weeks
earlier than expected, due to a power failure. Images were obtained whenever
the Sun elevation was sufficiently below the horizon. There were two time
periods during which no observations were made due to issues with the computer
system: 2008 May 19--29 and 2008 June 30 to 2008 July 15.

\section{RESULTS AND DISCUSSIONS}

\subsection{Sources of Sky Brightness}\label{sec-sky}

Night sky brightness is one of the key parameters for
qualifying a site for ground-based astronomical observation. Many
sources contribute to the sky background including scattering from
the Sun and the Moon, airglow, zodiacal light, aurorae, star light
and interstellar dust scattering, extragalactic light, and artificial
light contamination \citep{lei97,ben98a}. The contributions of these
sources to the sky brightness at Dome C have been reviewed by
\citet{ken06} and this is generally applicable to Dome A.

Artificial light pollution is essentially non-existent in Antarctica.
The main contribution to the sky background
is usually from the atmospheric scattering of the light of the Sun and the
Moon. \citet{ben98a} showed that zodiacal light (i.e., sunlight
scattered by interplanetary dust) may contribute up to half the
intrinsic sky brightness when the Sun and the Moon are down. However,
this is not an issue for CSTAR observations since
the field that we are observing is centered on the South Celestial Pole,
with ecliptic longitude $\lambda$
of $270^\circ$ and latitude $\beta$ $-66.^\circ6$, and is
far from the region affected by zodiacal light. On the other hand, we
expect an increased contribution from the Milky Way, since the Galactic
latitude $b$ of the South Celestial Pole is $-27^\circ$, relatively close to
the Galactic plane. We quantify this effect in Section \ref{gal}.

Another important source of light is from the aurorae, which are common
in polar regions. Aurorae are generated as electrons from the solar
wind which are trapped in the upper atmosphere by the magnetic field of the Earth.
The intensity and frequency of occurrence of aurorae
depend upon the solar activity. From satellite images and
ground-based observations, aurorae are known to be concentrated into
an oval-shaped annulus centered on the geomagnetic south pole. As
noted by \citet{dem05}, the more intense the solar activity, the
larger the extent of the oval. Domes A and C
both lie within the ``hole'' in the center of this oval-shaped region
and so are less seriously affected by aurorae than sites lying on the
oval itself. In 2008, solar activity was close to its minimum level
so that, perhaps paradoxically, Dome A would have experienced more aurorae
than in an average year. The maximum of solar cycle 24 is predicted
to appear in 2012 December \citep{kil09}, so the number and
intensity of aurorae appearing above Dome A should then be at a
minimum.

To better understand the effect of the Sun, the Moon, and cloud on the
sky brightness, we modeled the contributions of the Sun and the Moon
light without the influence of clouds in Section \ref{sun} and \ref{moon}.
After removing these two sources of sky light, we consider the relationship between the
residual sky background and atmospheric transparency (see Section \ref{ext}).
The remaining variable component of the sky brightness can then
be attributed to aurorae (see Section \ref{aur}).

\subsection{CSTAR Measurements of Sky Brightness}

The catalogs processed by the CSTAR pipeline contain the sky
background values in analog-to-digital converter counts (ADU). The
instrumental magnitude, $m$, in the CSTAR photometric system is defined
as
\begin{displaymath}
m_{\rm CSTAR} = -2.5 \textrm{log} (\textrm{ADU}) +25.
\end{displaymath}
\citet{zho10b} used the USNO-B1 (U.S. Naval Observatory) star
catalog to flux calibrate the CSTAR images and derived a
calibration constant of
$\overline{i_{\rm CSTAR}-i_{\rm USNO}}=4.16\pm0.12$.
SDSS $i$-band magnitudes are very
close to AB magnitudes \citep{boh01}, so  we can transform our
$i$-band magnitudes into a top-of-the-atmosphere flux. The
magnitudes in the AB system \citep{oke83} are defined as follows:
\begin{displaymath}
 m = -2.5 \textrm{log} F_\nu - 48.6,
\end{displaymath}
where $m$ is the AB magnitude and $F_\nu$ is the flux of the specified
wavelength band in the units of $\rm erg\ s^{-1}cm^{-2}
Hz^{-1}$. Then we can convert from CSTAR raw ADU counts (normalized to
a 20s exposure) to an $i$-band sky flux per square arcsec using
\begin{displaymath}
F = \textrm{ADU} \times 10^{-27.798}/15^2 = 0.708\times10^{-30}
\textrm{ADU},
\end{displaymath}
where we have assumed that the typical
atmospheric extinction in clear photometric nights is about 0.05
mag/airmass at $i$, taken from the observational
results of the Carlsberg Meridian Telescope (CMT) at La Palma, and $15^2$ is the
pixel area in square arcseconds.

The surface fluxes in the rest of this paper are expressed in units
of $10^{-30}\rm ergs\ s^{-1}cm^{-2} Hz^{-1}arcsec^{-2}$.

\begin{figure}
\epsscale{1.0} \plotone{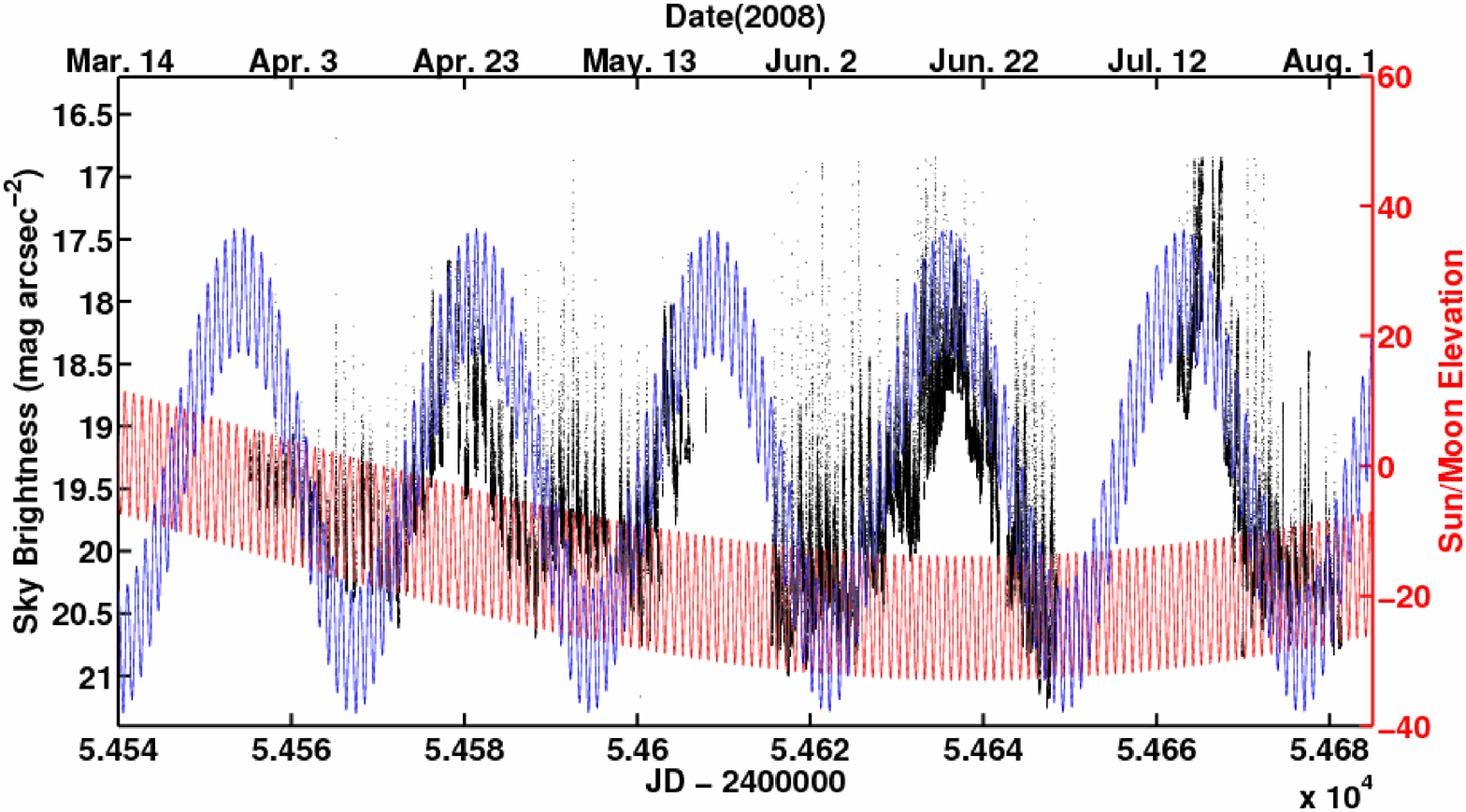} \caption{Sky brightness
in $i$ band (black dots). The
red and blue curves are the elevations of the Sun and the Moon,
respectively. \label{fig1}}
\end{figure}

Figure \ref{fig1} shows the sky brightness variations throughout the
whole observing period of 2008. A diurnal
cycle is evident even in midwinter. At such far-southerly latitudes as Dome A,
the Moon is always fairly full whenever
it is above the horizon from April to August, creating a
strong correlation between lunar elevation (including
the correction for parallax) and sky brightness.

\begin{figure}
\centering
\epsscale{0.5}
\includegraphics[width=0.5\textwidth]{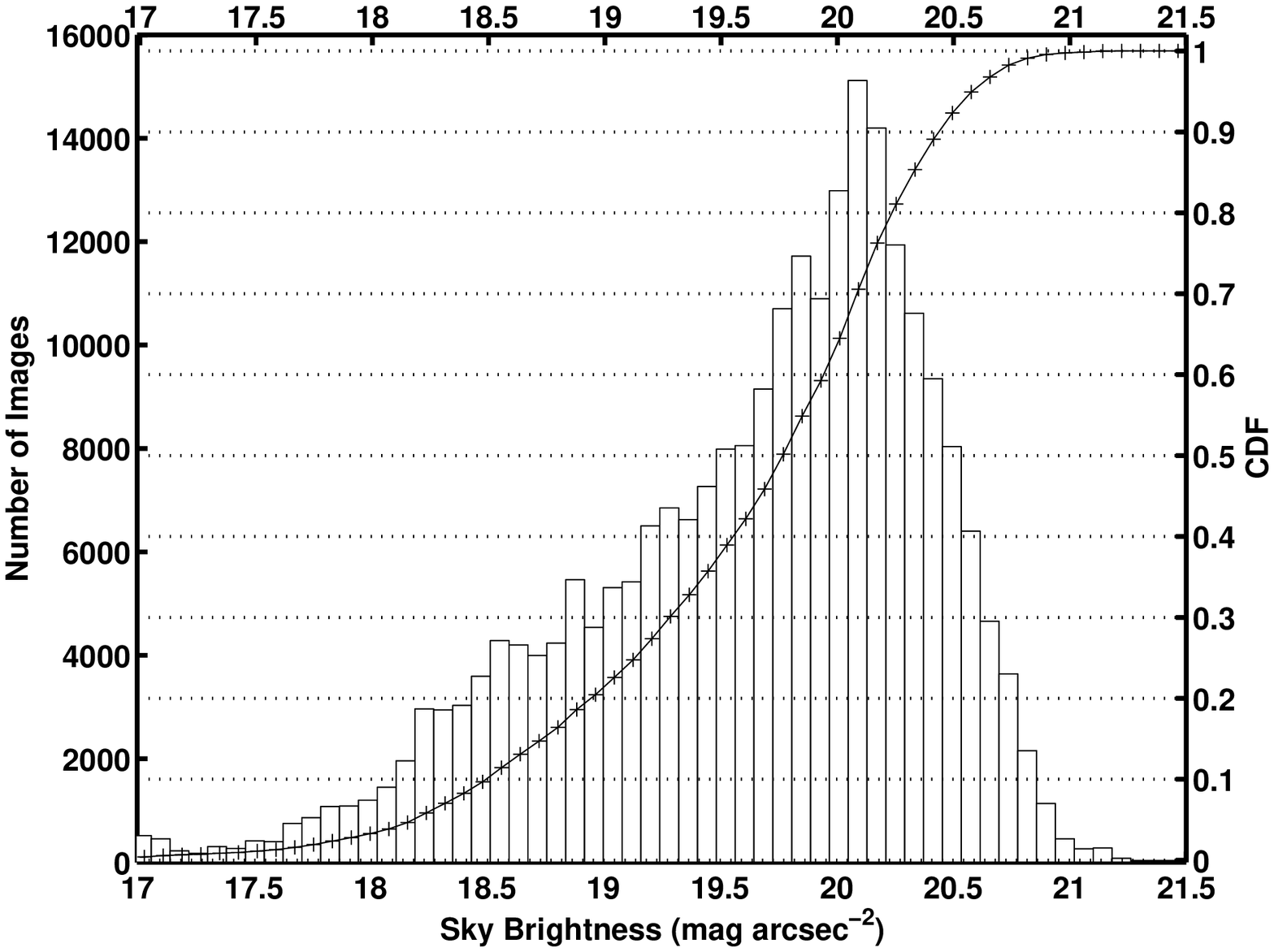}
\hfil
\epsscale{0.5}
\includegraphics[width=0.45\textwidth]{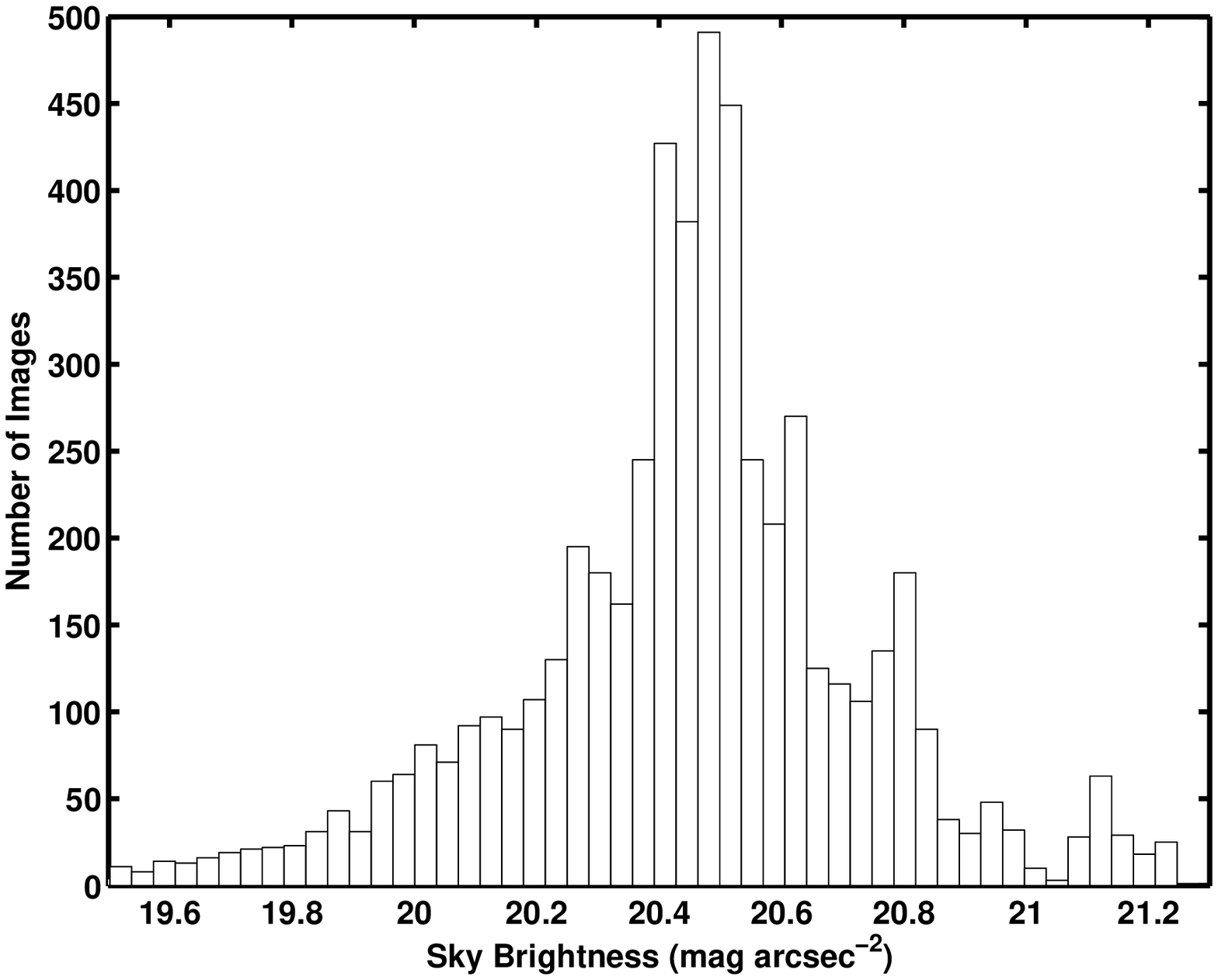}
\caption{Left panel shows the histogram and the cumulative distribution function (CDF) of the $i$-band sky brightness
distribution at Dome A during 2008. Right panel is the same information for the subset of
images taken on moonless clear nights in
2008 June.\label{fig2}}
\end{figure}

Figure \ref{fig2} shows histograms of the sky brightness
distribution at Dome A. The most probable sky brightness, across
all lunar phases (left panel of the Figure), is found to be 20.1
mag arcsec$^{-2}$ and the median is 19.8 mag arcsec$^{-2}$.
About 80\% of the images have sky
backgrounds darker than 19 mag arcsec$^{-2}$.

The right panel of Figure \ref{fig2} shows the sky brightness
distribution of images taken only on moonless and clear nights during 2008
June. In these images, the Sun and the Moon elevation is below
$-18^\circ$ and there is no cloud. We find that the median sky
brightness is 20.5 mag arcsec$^{-2}$. When it is dark but
cloudy, the sky brightness increases to about 20.2 mag arcsec$^{-2}$. When
the Moon is full and it is clear, the brightness is around 19.1 mag
arcsec$^{-2}$. When the Moon is full and there are clouds, the average
sky background is 17.9 mag arcsec$^{-2}$.

\citet{mol10} have summarized the night sky brightness
in $UBVRI$ measured at various observatories during dark time.
We used the transformation equations
\citep{jor06} between SDSS magnitudes and $UBVRI$ to
compare our $i$-band measurements with the sky backgrounds of La Palma, Cerro Tololo,
and Paranal with Dome A. These transformations, which are appropriate for stars,
are slightly different from those that should be used for the sky background because
of the presence of strong emission lines. We calculate the following
median $i$-band sky brightnesses: 20.10
mag arcsec$^{-2}$ at La Palma (at sunspot minimum; \citet{ben98b}), 19.93 at Paranal
(at sunspot maximum; \citet{pat03}), 20.07 at Cerro Tololo (at sunspot minimum; \citet{wal87,wal88}) and
19.57 at Calar Alto (at sunspot maximum; \citet{san07}).
The CSTAR measurements reported here, with a clear dark sky brightness
of 20.5 mag arcsec$^{-2}$, support the tentative conclusion that
under moonless clear conditions, Dome A has a darker sky background than
the above astronomical sites, even allowing for calibration and
transformation uncertainties of up to several tenths of a magnitude.

\subsection{Correction for Galactic Background}\label{gal}

To compare our observations of the sky background at the South
Celestial Pole with observations taken well away from the Galactic
Plane we need to estimate the contribution from diffuse Galactic
light. The Pioneer 10 and 11 spacecraft made observations of the total
Galactic plus extragalactic sky background in blue (395--485nm) and
red (590--690nm), from beyond the asteroid belt. The results for
selected fields, including the South Celestial Pole, are given in
\citet{tol87}. The Pioneer values have a resolution of about 2$^\circ$
on the sky and include the light from stars fainter than about
$V\sim8$. We extrapolate the Pioneer red data into the $i$ band by
using the ratio of line-of-sight integrated starlight \citep{mat80}
and the ratio of line-of-sight extinction at the different wavelengths. This
leads to an estimate of $26\pm8\times10^{-10}{\rm erg\,\,cm}^{-2}{\rm
  s}^{-1}{\rm sr}^{-1}{\rm nm}^{-1}$, which is a 23.75 mag
contribution to our median of 20.5 mag (i.e. $\sim5\%$ of the total sky
background). In other words, we would expect the median sky brightness
in regions well away from the Galactic plane to be a further
$\sim 60$~mmag fainter.

\subsection{Transparency Variations}

Starlight is attenuated as it passes through Earth's atmosphere
due to molecular absorption and scattering from molecules and
aerosols. Ideally, there would be stable, dry air above the site
and, in the absence of clouds, the observed flux of each
non-variable star would show little change. Normally, it would be
possible to derive the absolute extinction for the site by observing
the same star at two very different airmasses or by observing stars
of known brightness that happen to be at two different airmasses.
However, in the case of CSTAR, with only a 4.$^\circ$5 field, neither technique is practical. We thus
confine our discussion to {\it variations\/} in transparency, with
the absolute value of the extinction yet to be determined, although presumably
it is low at Dome A due to the altitude of the site and the low aerosol content
of the atmosphere.

To assess the variations in transparency, we compared the observed
brightness of an ensemble of bright, unsaturated, stars to the
brightness of these stars on a reference image, chosen as one
containing the greatest number of
detected point sources. All images were scaled to the same exposure time, 20 s.

\begin{figure}
\epsscale{1.0} \plotone{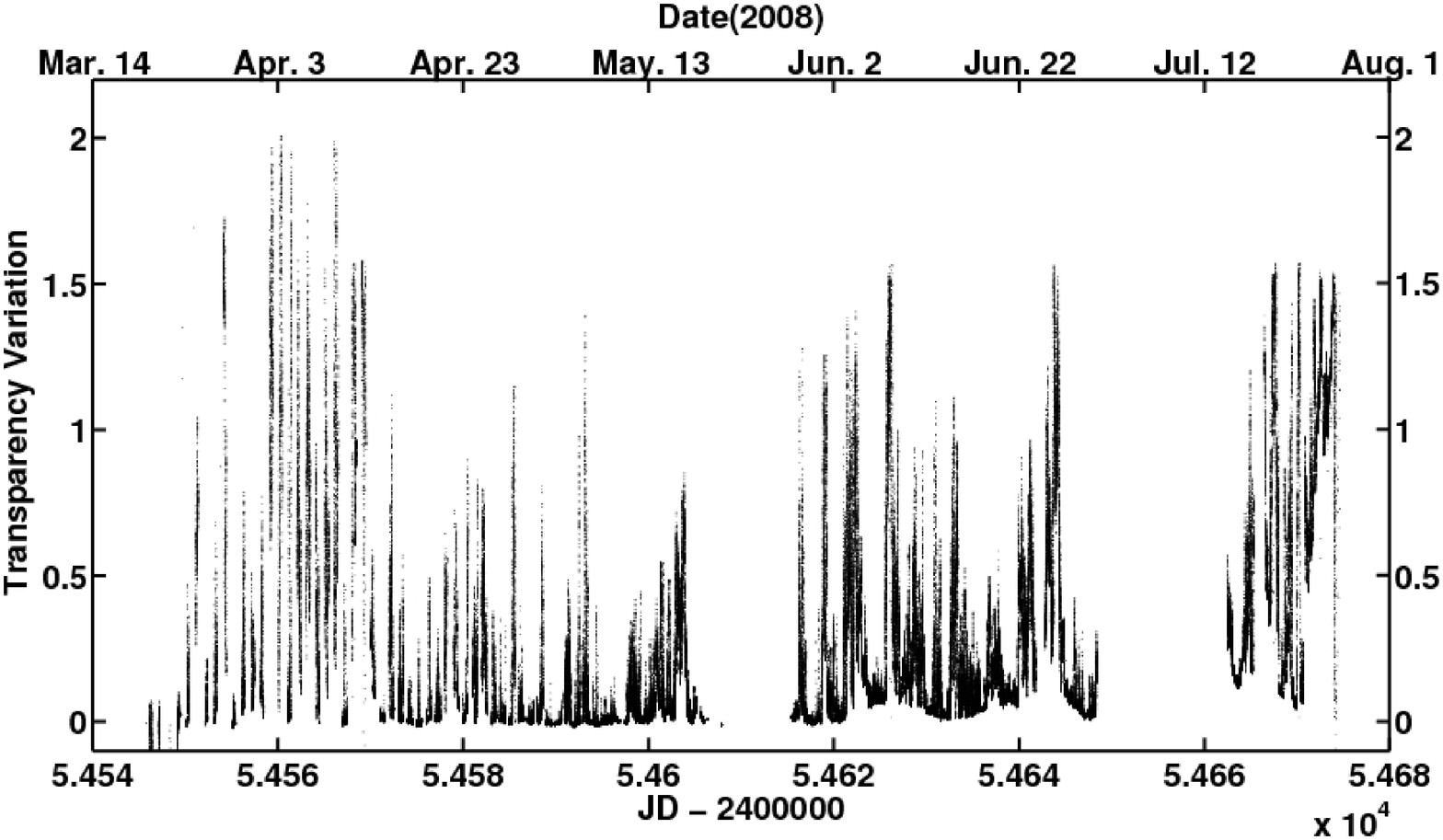} \caption{$i$-Band sky transparency variation (in magnitudes) vs.
Julian Date (JD) for all the CSTAR images during 2008.\label{fig3}}.

\end{figure}

\begin{figure}
\epsscale{1.0} \plotone{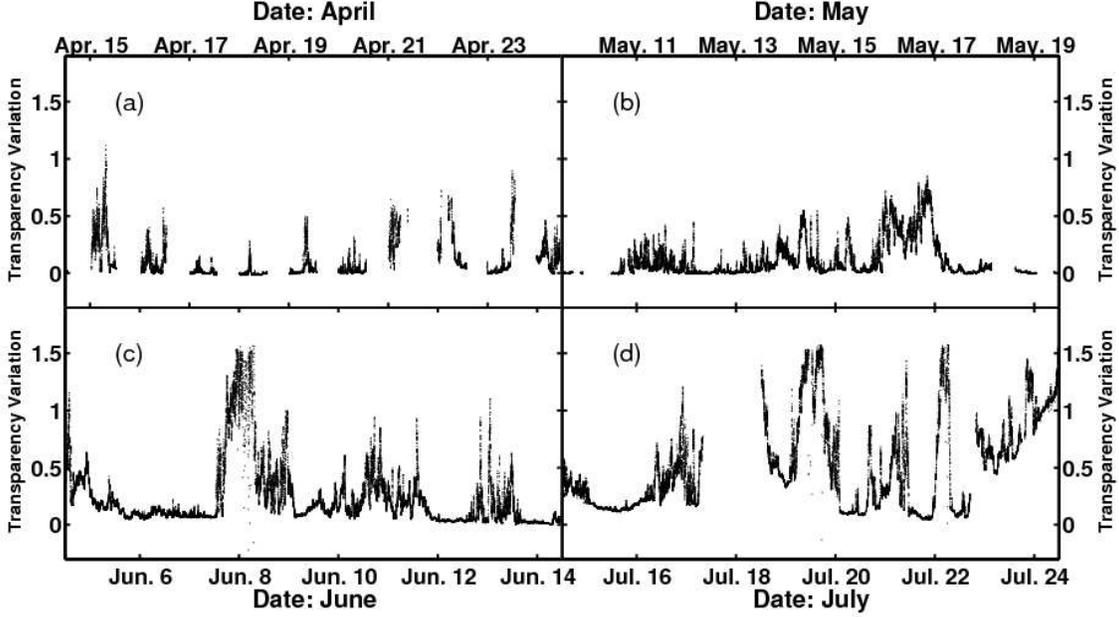} \caption{(a) $i$-Band sky transparency variations (in magnitudes)
from April 15 to April 25. The gaps are periods when sunlight made the sky too bright to
observe. (b) May 10 to May 20, (c) June
5 to June 15, and (d) July 15 to July
25.\label{fig4}}
\end{figure}

Figure \ref{fig3} presents the time variation of
atmospheric transparency during the periods of darkness throughout 2008, while
Figure \ref{fig4} shows an expanded version covering 10 day periods
in each month from April to July.
From April 15 to June 2, the transparency was relatively stable with many
days showing high transparency. In June and July, the sky
reaches its darkest due to the low elevation of the Sun;
the transparency became a little worse in July. The median
values of the excess extinction, above that of our nominally clear reference image, for the months from April to July are 0.056, 0.016,
0.127, and 0.49, respectively.

\begin{figure}
\epsscale{1.0} \plotone{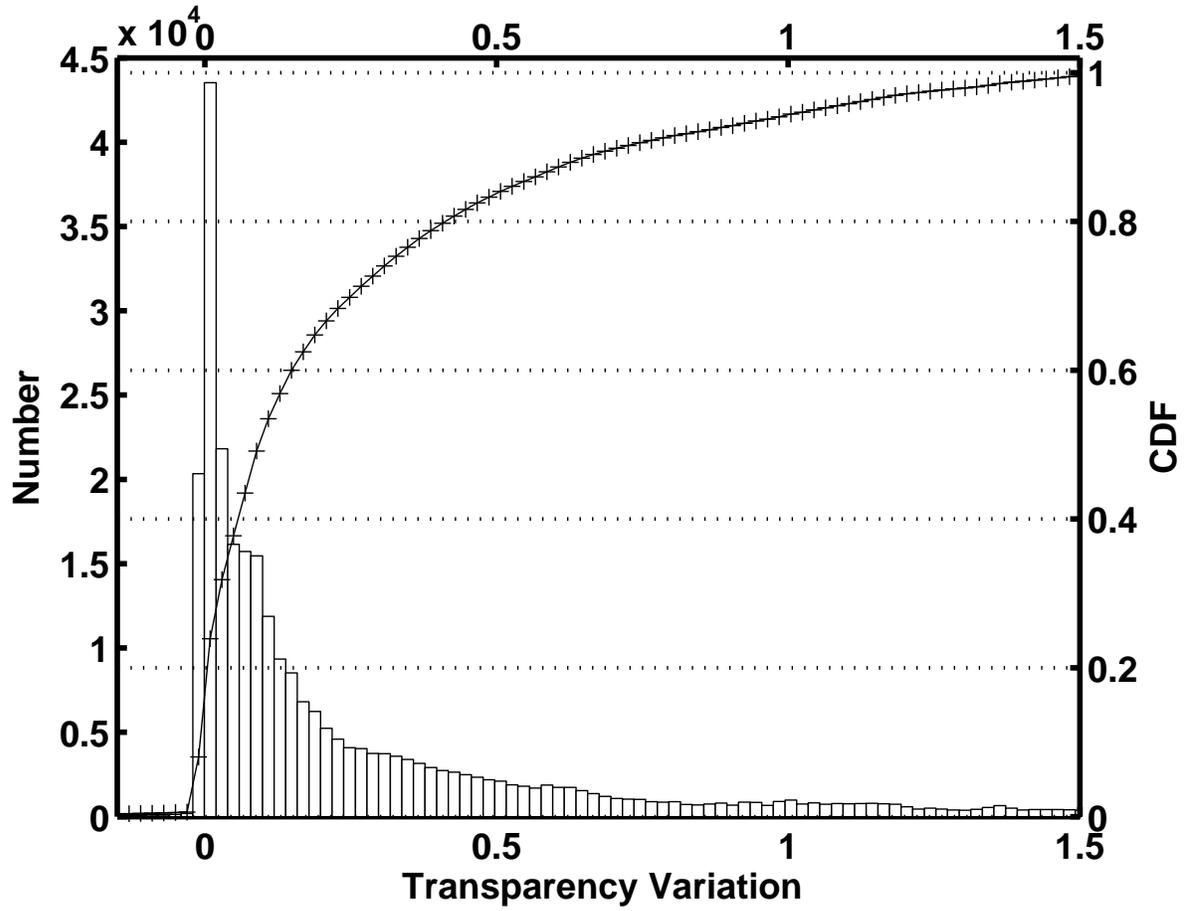} \caption{Histogram of the number of images
with the given transparency variation (in magnitudes) relative to the reference image.
The curved line marked `+' is the CDF. \label{fig5}}
\end{figure}

Figure \ref{fig5} shows the distribution of the transparency. Most of
the images have only a small increase in transparency relative to the
reference image: 90\% of the images have less than 0.7 mag of
transparency, 80\% have less than
0.4 mag, and more than half have less than 0.1 mag.
\citet{ive07} show that in conditions of up to 1 mag of reduced transparency,
photometry is limited by photon noise, not by variations in the cloud
extinction itself. Therefore, conditions were suitable for accurate photometry
for $>90$\% of the time during 2008 at Dome A.

\subsection{Percentage Cloud Cover}
Here, we use our observed relative transparency variations to infer
the distribution of the optical thicknesses of cloud at Dome A during 2008.
We note that any reduction in transparency could be due to combinations of cloud,
the atmosphere, possible ice formation on
the entrance pupil of the telescopes, and the exhaust plume produced by
the diesel generators (about 45 m away from CSTAR). Our results
therefore represent upper limits to the cloud. We expect window frosting to be
negligible, due to the design of the front window of the
telescope which is heated all the time while observing.
Table \ref{tab2} lists the percentages of images in various
extinction ranges---chosen to correspond to the flux percentage of stars
relative to the flux in the reference image as shown in the first column.
About $51\%$ of the images were obtained in excellent conditions
with excess extinction from any cause of less than 0.11 mag.

For a rough comparison with another observatory, Table \ref{tab3}
shows the approximate fraction of cloud of various thicknesses at Mauna Kea.
The data for the table come from the Gemini
Observatory\footnote{{\url{http://www.gemini.edu/sciops/telescopes-and-sites/observing-condition-constraints}}}
and are based on nightly logs from the United Kingdom Infra-red Telescope
over a 10 year period. The relative transmission
levels Gemini provides are in $V$-band magnitudes (these levels are
subjectively assumed by them for the purpose of integration time calculation).
We use the calculated atmospheric extinction \citep{kin85} to convert these $V$-band extinction
in clear weather to $i$ band by simply applying the differential
extinction between these two passbands ($V - i = 0.07$ from the
observational results of the CMT for La
Palma on Web site of \url{http://www.ast.cam.ac.uk/~dwe/SRF/camc_extinction.html}).
Here, $V - i = 0.07$ is just like a systematic zero point to transform
the cloud extinction ranges of different cloud categories between these two
bands, although it is unclear about how the cloud extinction varies
in different passbands and from site to site. Table \ref{tab3} shows that the cloud cover
at Dome A falls into the ``cloudy'' or worse category for a mere 2\% of
the time, compared with 30\% at Mauna Kea. There are patchy clouds at
Dome A for about 31\% of the total usable time, most of which are cirrus passing by
as verified by wide-field cameras on the top of PLATO when the Moon is up.
Overall there is a greater fraction of photometric conditions at Dome A
compared to Mauna Kea. However, these conclusions are tentative due to the
limited sky field of CSTAR, different photometric systems, and the fact
that the Dome A data are from only 5 months during 2008.

\subsection{Absolute Flux from the Sun} \label{sun}

The Sun and the Moon have of course a strong effect on the sky
background, especially when they are above the horizon.
\citet{liu03} modeled the relationship between
the sky brightness and the phase and elevation of the Moon. The way
in which the Sun and the Moon affect the sky background should be
formally analogous except for the additional complication that the
Moon exhibits phases.

\begin{figure}
\epsscale{1.0} \plotone{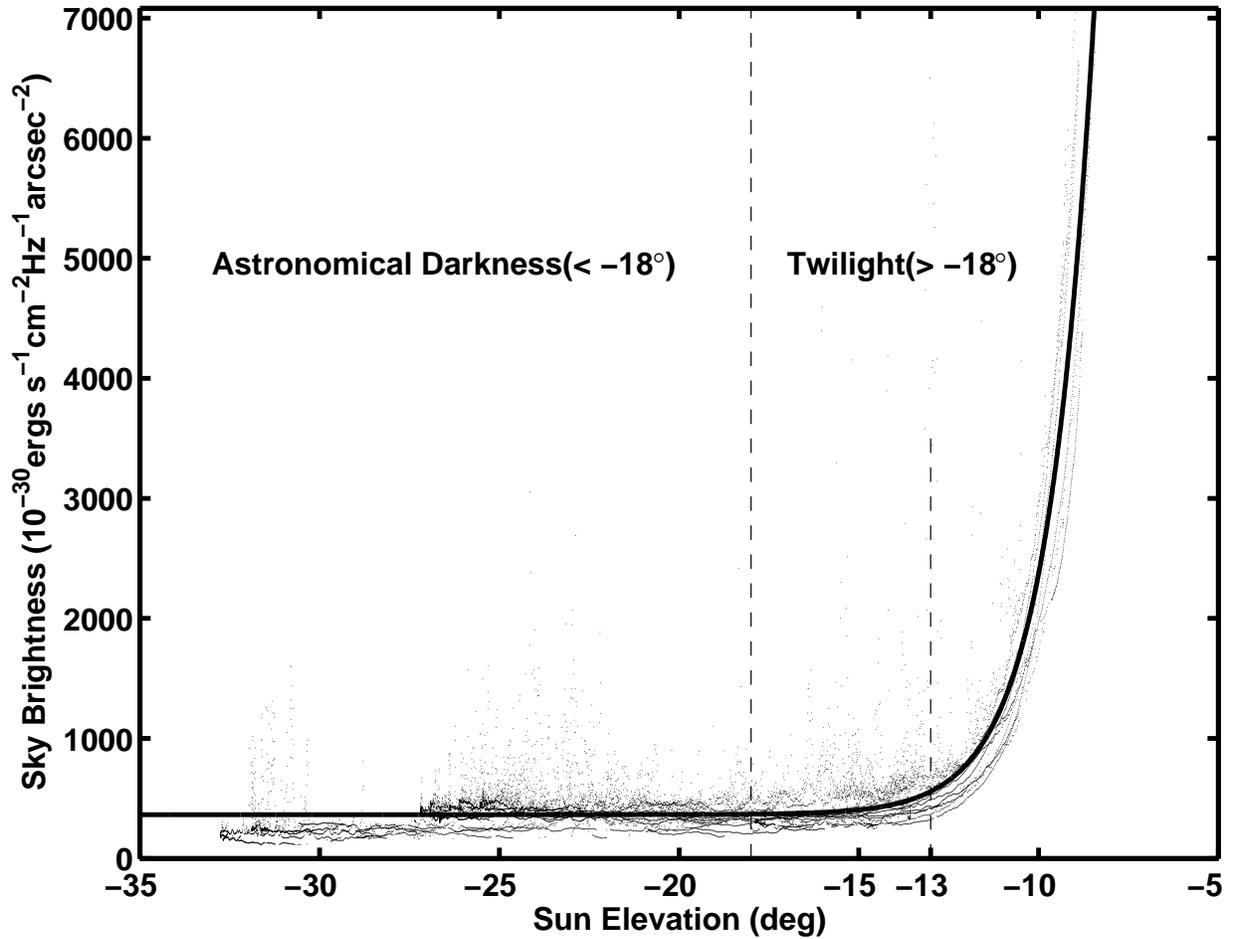} \caption{Relation between the sky
brightness and the elevation of the Sun. The data points are those images
that were taken in good conditions with negligible contribution from
the Moon. The curved line is the fit from Equation (\ref{sunfit}).
The dashed line at $-18^\circ$ gives the dividing elevation where the
astronomical darkness is defined to begin. The dashed line at $-13^\circ$ indicates a
more suitable cutoff for $i$-band dark time at Dome A. \label{fig6}}
\end{figure}

Referring to the model of \citet{liu03} in $V$ band, we fit the sky background flux as a
function of the solar elevation with the same formula for our $i$-band data,
\begin{equation}
 F_{\rm Sun} = a 10^{b\theta} + c, \label{sunmod}
\end{equation}
where $F_{\rm Sun}$ is the sky flux (assuming only a solar contribution and a background),
$\theta$ is the elevation of the Sun, and $a$, $b$ and $c$ are constants.
We fitted for the constants with a nonlinear least-squares method, using the images with good
transparency and no contribution from the Moon. The result is
\begin{equation}
F_{\rm Sun} = 4.906\times10^6\times10^{0.339\theta} +370 \label{sunfit}
\end{equation}
and is plotted against the observations in Figure \ref{fig6}.
We find that the sky is about as dark as it ever gets in $i$-band
once the Sun is below about 13$^\circ$ from the horizon.

\subsection{Astronomical Twilight and Hours of Dark Time}

Astronomical twilight is defined as when the Sun is between $12^\circ$ and $18^\circ$ below
the horizon. However, this definition is really only appropriate at
low altitude observatories and in the $V$ band.
At the high elevation of Dome A, and in $i$ band, the sky is as dark as it
ever gets when the solar elevation reaches about $-13^\circ$ (see Figure \ref{fig6}).
Table \ref{tab1} summarizes the numbers of hours when
the Sun is located below elevations of $-18^\circ$, $-13^\circ$, and $0^\circ$.
If we require only that the Sun is below the horizon, Dome A has more
``night time'' that any of the other sites. This is because Earth's
elliptical orbit around the Sun results in the total hours of night time
in a year increasing with distance from the north pole. However, for
``dark time'' to occur, we require the Sun to be at least a certain
elevation below the horizon (conventionally,  $-18^\circ$).
There is less dark time at Dome A compared to temperate-latitude observatories,
although the $-13^\circ$ solar elevation limit makes this difference less
dramatic than it would be at $-18^\circ$. While the total dark time is less,
the long periods of continuous darkness are an advantage for many observations.
Over $83\%$ images were obtained during 2008 when the Sun was $13^\circ$ below the horizon.
The discussion in the rest of this paper is based on these images.

\subsection{Absolute Flux from the Moon}\label{moon}
The model for the sky surface brightness caused by moonlight is more
complex. In our analysis, we ignore the variation in the lunar
brightness caused by the changes in the Moon$-$Earth distance.
Following \citet{liu03}, the apparent magnitude of the Moon can be
approximated by an empirical formula:
\begin{displaymath}
 V(R,\Phi) = 0.23 + 5 \textrm{log} R - 2.5 \textrm{log} P(\Phi),
\end{displaymath}
where $R$ is the Moon$-$Earth distance, and $\Phi$ is the Moon phase
angle, and $P(\Phi)$ is the function of the full Moon illuminance.
Then, using the same approach as for the Sun, the sky surface flux
contribution by the Moon, $F_{\textrm{moon}}$, can be expressed in
the form of Equation (\ref{sunmod}) multiplied by $P(\Phi)$. Then,
\begin{equation}
 F_{\textrm{moon}} = A P(\Phi)10^{B\Theta} + C, \label{moonmod}
\end{equation}
where $\Theta$ is the elevation of the Moon, and $A$, $B$, and $C$ are
constants. $C$ is the intrinsic background flux to be fitted. We applied
these equations to the $i$- band data. Although the model is somewhat crude,
it works well in the estimation of sky brightness illuminated by the Moon.
For a more refined but slightly more complicated sky brightness model, we can
refer to the study of \citet{kri91}.

\begin{figure}
\epsscale{1.0} \plotone{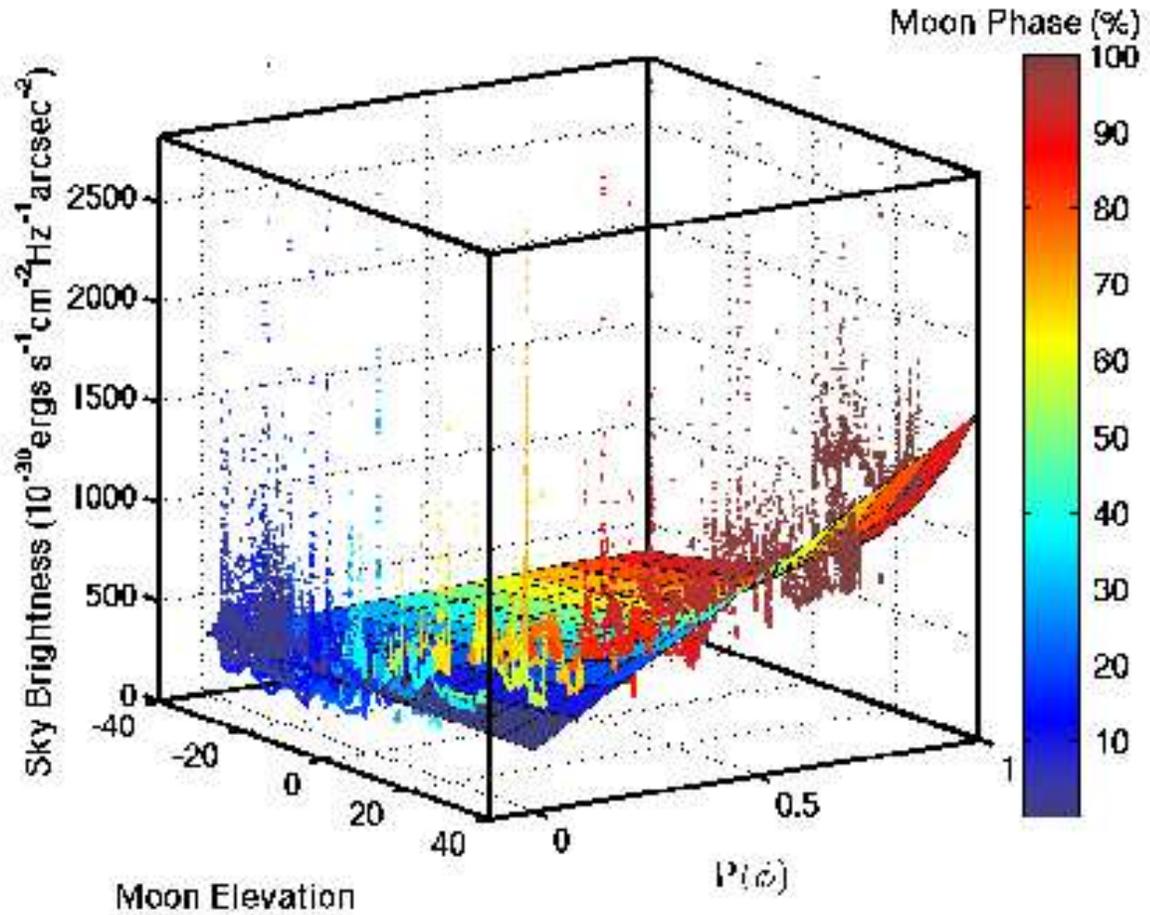} \caption{Relationship among the sky
brightness (with the solar contribution subtracted and under conditions of good relative transparency),
the lunar phase, and the elevation of the Moon. The curved and colored surface is
fitted by the model in the form of Equation (\ref{moonmod}). The color
bar is related to the Moon phase in percentage, which is the
illuminated fraction of the full Moon disk. The outliers may be
aurorae. \label{fig7}}
\end{figure}

Figure \ref{fig7} displays the Moon contribution to sky surface
brightness as the function of the Moon phase and elevation. The sky
brightness in the figure is the Sun-corrected sky surface flux
without any contribution from clouds. The colored curved surface is
the fitted model described by
\begin{equation}
F_{\textrm{moon}} = 300.576P(\Phi)10^{0.0162\Theta} + 306.813. \label{moonfit}
\end{equation}
We can see that the sky was brightest when the Moon was almost full.
The outliers should be the points when aurorae were
present.

\begin{figure}
\epsscale{1.0} \plotone{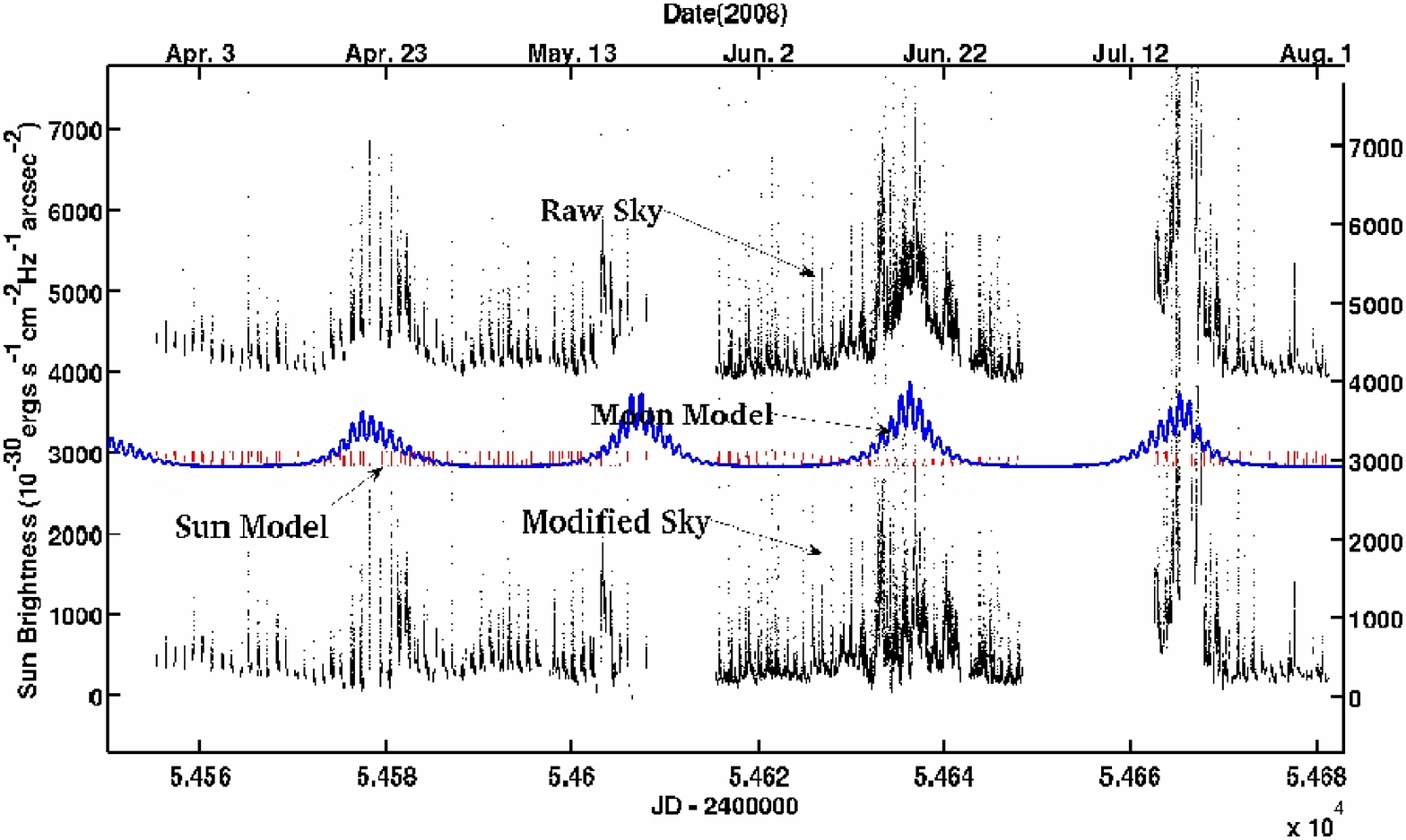} \caption{Sky brightness before and
after correction (subtracting the Sun and Moon contributions) and
the contributions of the Sun and the Moon in the fitted models.
The model of the Moon is displayed in continuous JD, but the model
of the Sun is plotted at the time samples those same
as the sky brightness. The original sky brightness and the model
fluxes of the Sun and the Moon are shifted vertically from their proper
values.\label{fig8}}
\end{figure}

\begin{figure}
\epsscale{1.0} \plotone{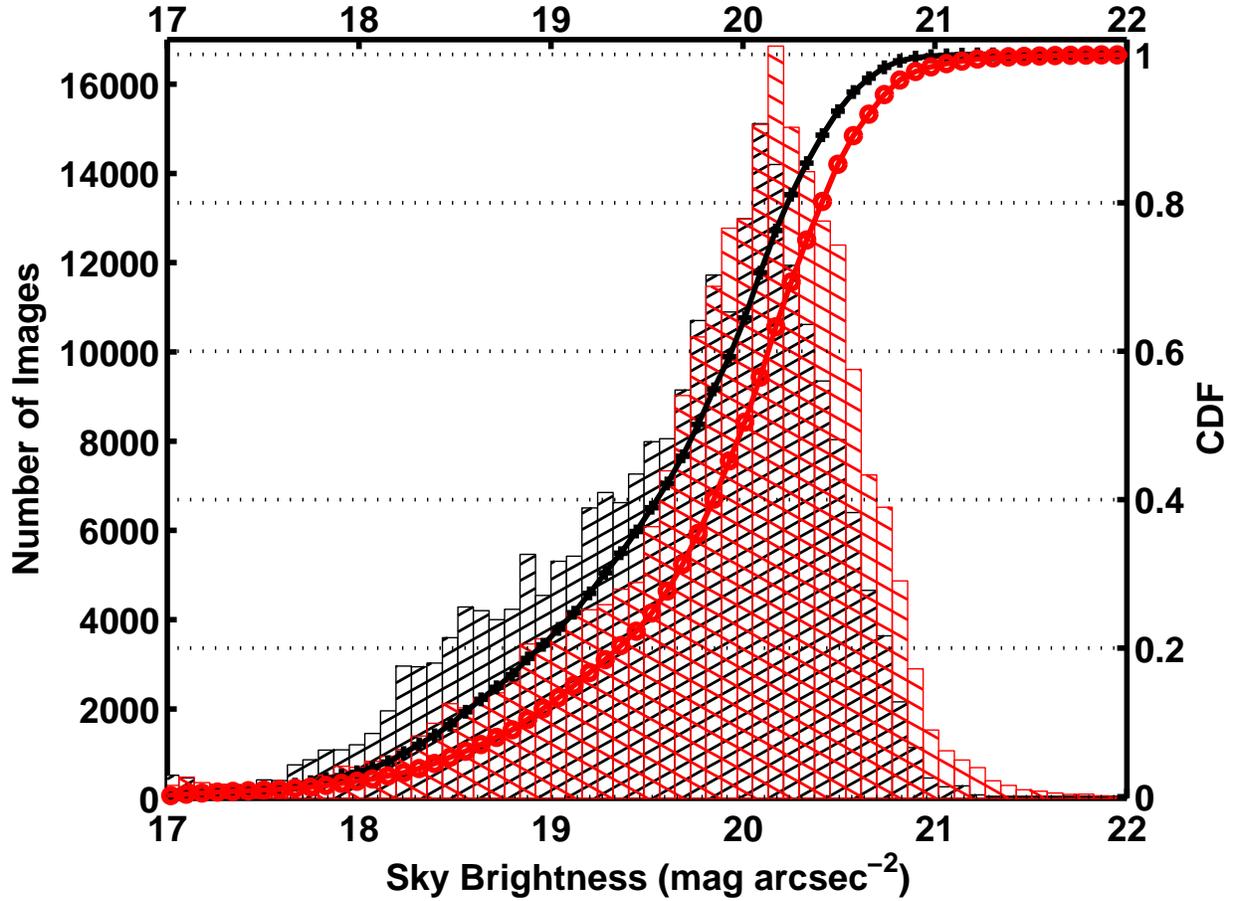} \caption{Histograms of the sky
brightness distributions of both the original data and the modified one
(subtracting the Sun and Moon contributions). The area patched with black
slashes and the curved line with markers '+' represent the raw sky brightness
and its CDF. The area with red backslashes and the curved line with circles are the
modified sky background and its CDF. \label{fig9}}
\end{figure}

We compare the modified sky background (removing both the solar and
lunar contributions) with the raw one (before removing the solar
contribution) in Figure \ref{fig8}. The modified sky background
becomes flat wherever the Moon is up. Figure \ref{fig9} compares the histograms of the raw sky
brightness in unit of mag arcsec$^{-2}$ (the left panel shown in
Figure \ref{fig2}) with the modified sky background. The histogram
with black slashes represents the original data and the one patched
with red backslashes is for the modified sky brightness. The median
sky brightness becomes 20.04 mag arcsec$^{-2}$ after being corrected. About
80\% of the images are fainter than 19.35 mag arcsec$^{-2}$.

\subsection{Correlation Between Sky Brightness and transparency variation} \label{ext}
\begin{figure}
\epsscale{1.0} \plotone{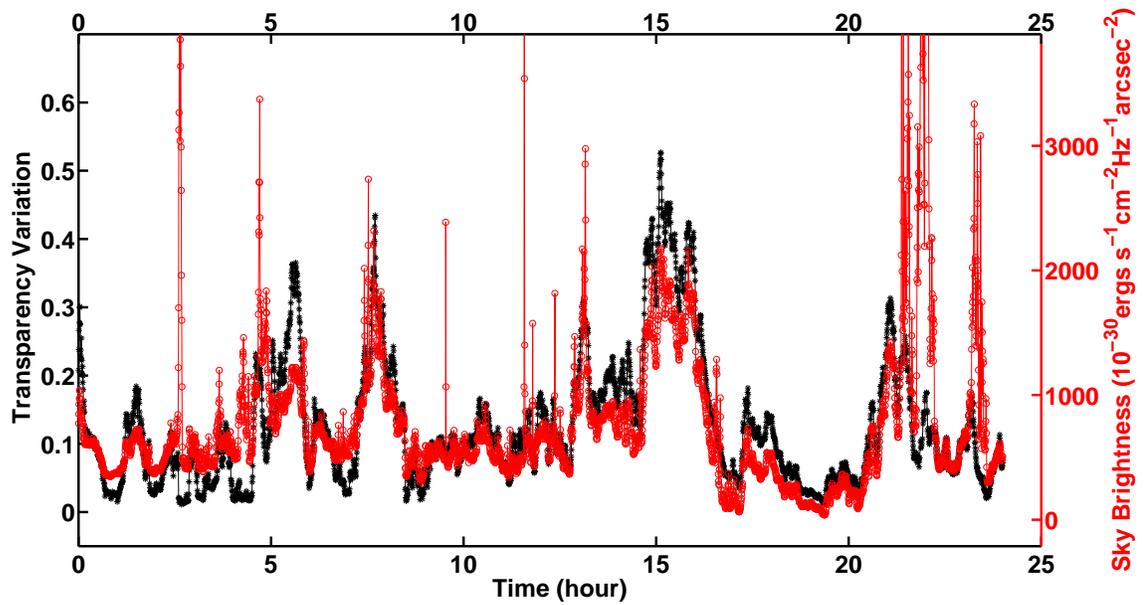} \caption{Variations of the sky
brightness and transparency variation (excess extinction in magnitudes) for 1 day in 2008. The black asterisks are the
excess extinction and the red circles are the sky background. The $x$-axis is
the time in hours elapsing from 2008 June 16 00:00 UTC.
\label{fig10}}
\end{figure}

\begin{figure}
\epsscale{1.0} \plotone{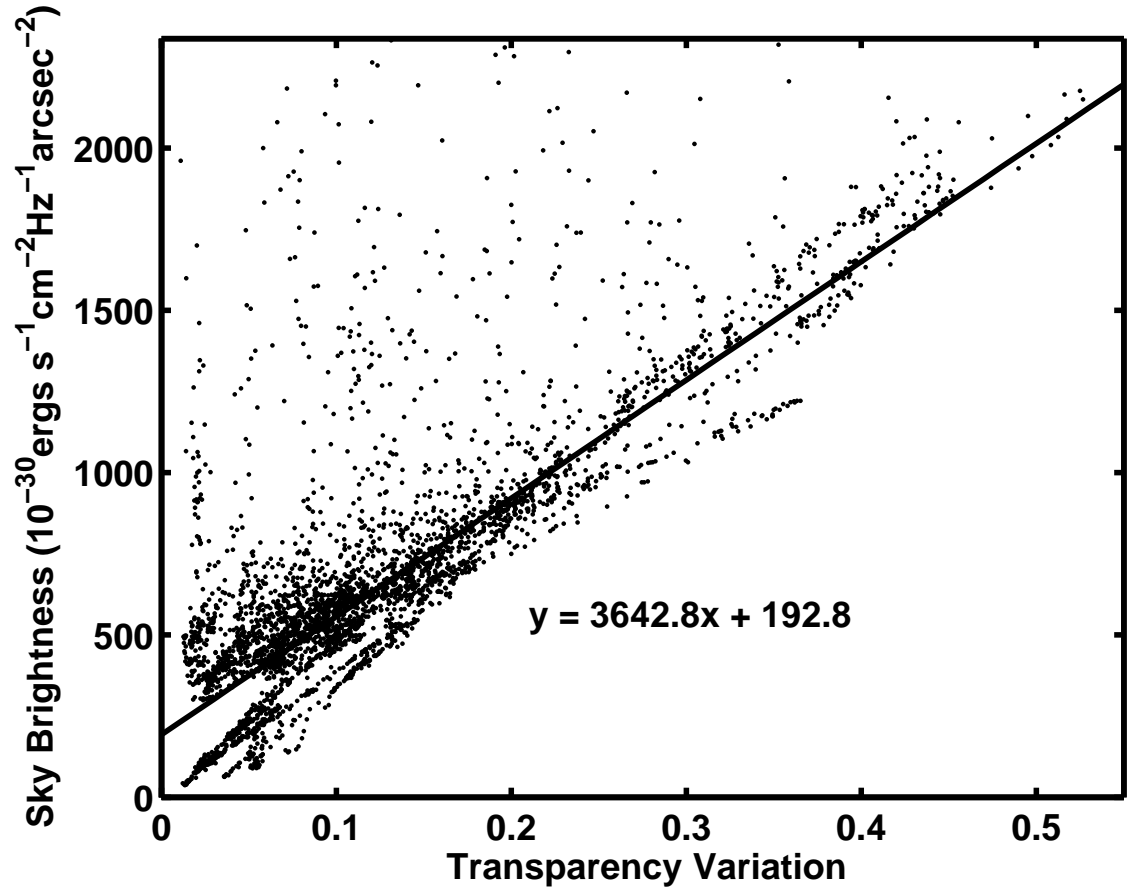} \caption{Correlation of the
corrected sky brightness and transparency variation (in magnitudes). The data points are
observations taken on 2008 June 16, when the average
elevations of the Sun and the Moon are $-23.^\circ17$ and
$25.^\circ12$ respectively, and the lunar phase was $\sim95$\% ($\Phi = 25.8$, $P(\Phi) = 0.54$). The
straight line is the fitted relation with the equation shown. \label{fig11}}
\end{figure}

The sky will be brighter when there is scattering of the Sun and the
Moon light by clouds. In order to study this effect, we choose a
day, June 16, when the Moon was nearly full, the Sun was far below
the horizon, and the transparency was varying. Figure \ref{fig10} shows
the variations of the sky brightness and relative transparency in this day. We
can see that they are strongly correlative, and the plot of the sky
brightness against transparency in Figure \ref{fig11} confirms that
the relationship is linear (again, outliers are the data polluted by aurorae).
The line of best fit and the derived linear equation are also shown in the figure.

The sky brightness is proportional to both the extinction (here, relative transparency)
and the total illuminance of the Sun and the Moon. The larger the extinction, the
brighter the sky background for a given total illuminance. The greater
the total illuminance, the brighter the sky background for a given
extinction.

We construct a model describing the sky brightness as a function of
the total illuminance and transparency variation. When the relative transparency is negligible,
the sky background should be the intrinsic sky background (a constant value),
resulting in the model not containing an individual term of the Sun and the Moon.
At new Moon or when the Moon is low enough below the horizon and the
elevation of the Sun is low, the sky brightness should be the sum of
the intrinsic background and cloud contribution. This requires a
single relative transparency term. Then:
\begin{displaymath}
 F_{\textrm{sky}} = a (F_{\textrm{sun}}+F_{\textrm{moon}}) E + bE + c,
\end{displaymath}
where $F_{\textrm{sky}}$ is the corrected flux of the sky light,
$F_{\textrm{sun}}$ and $F_{\textrm{moon}}$ are the model fluxes of
the Sun and the Moon given by Equations (\ref{sunfit}) and (\ref{moonfit})
(without constant terms), $E$ is the relative transparency and, $a$, $b$ and
$c$ are constants. $c$ denotes the fitted intrinsic background and
$b$ should be less than zero, indicating that the sky is dimmed by
the clouds when it is a moonless night and the Sun is very low. The result,
which gives an excellent fit to the data, is
\begin{equation}
 F_{\textrm{sky}} = 5.4751(F_{\textrm{sun}}+F_{\textrm{moon}})E -
 93.98E + 291.96. \label{extfit}
\end{equation}

\subsection{Percentage of Images Affected by Aurorae} \label{aur}
\begin{figure}
\epsscale{1.0} \plotone{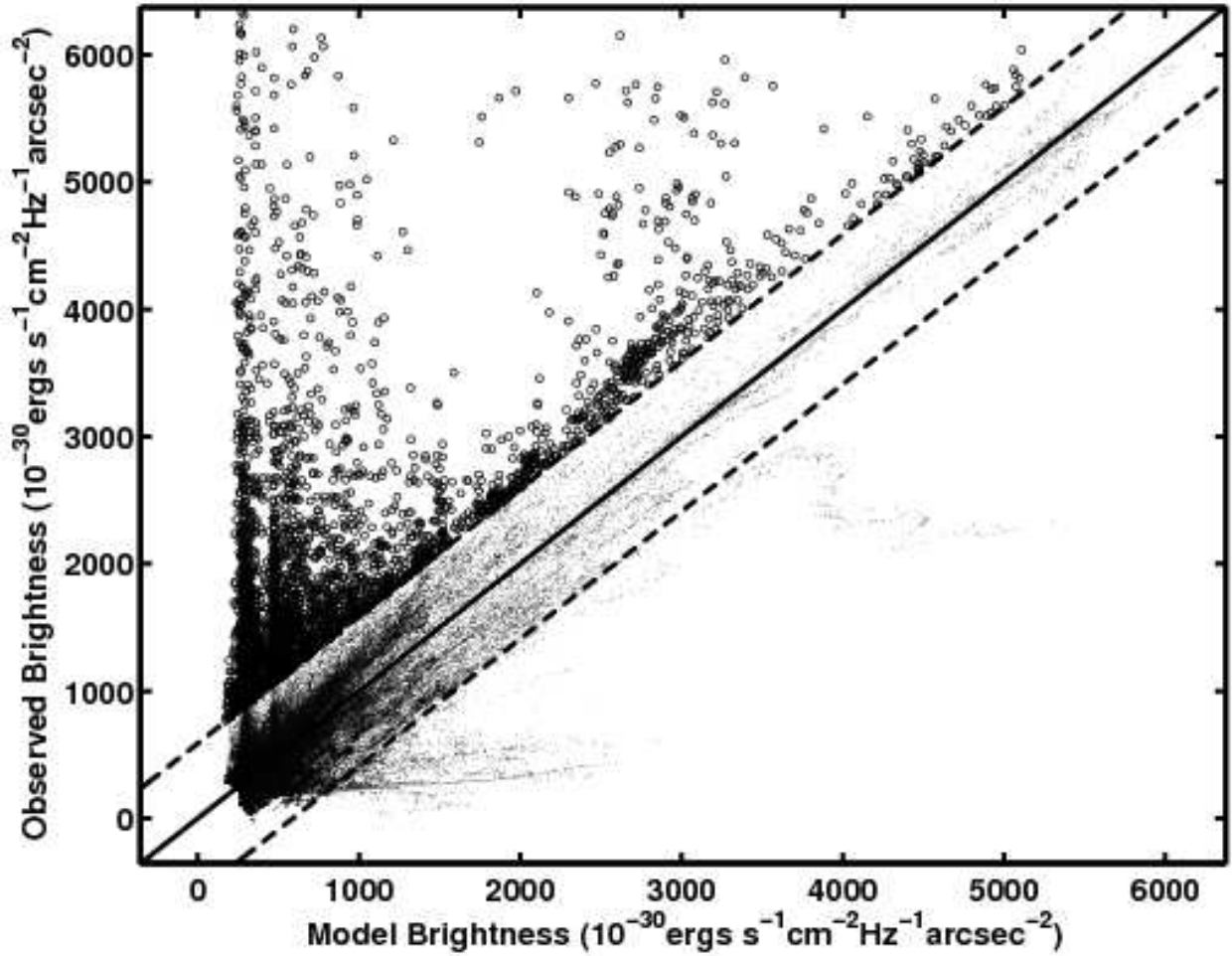} \caption{Observed sky brightness
corrected for the Sun and Moon contributions vs. the values
predicted by the fitted model from Equation (\ref{extfit}).
The diagonal denotes where the observations and model agree exactly.
The two dashed lines are 3$\sigma$ above and below this level.
The circles show the observed data points, presumably affected by aurorae, with
brightness brighter than 3$\sigma$ above the model. The transparency
variations are in magnitudes.
\label{fig12}}
\end{figure}

For our SDSS $i$ band (effective wavelength: 780nm; bandwidth: 160nm \citep{zho10a}), the
emissions of aurorae within this band mainly contain the lines from
molecular nitrogen and molecular oxygen in the lower atmosphere and those
from atomic oxygen in the higher atmosphere. The typical green$-$yellow auroral
emission lines are dominated by the atomic oxygen, lying outside the $i$ band \citep{dem05}.
To estimate the percentage of images affected by aurorae, we use the model of
sky brightness as described in Equation (\ref{extfit}).
Figure \ref{fig12} shows how well the model fits the data and
identifies the outlying points (implying the presence of aurorae) as circles.
We define an outlier---and hence the image affected by an aurora---as a point
that is more than 3$\sigma$ brighter than the model would predict
(although it is also possible that a small fraction of those images are affected by airglow).

\begin{figure}
\epsscale{1.0}\plotone{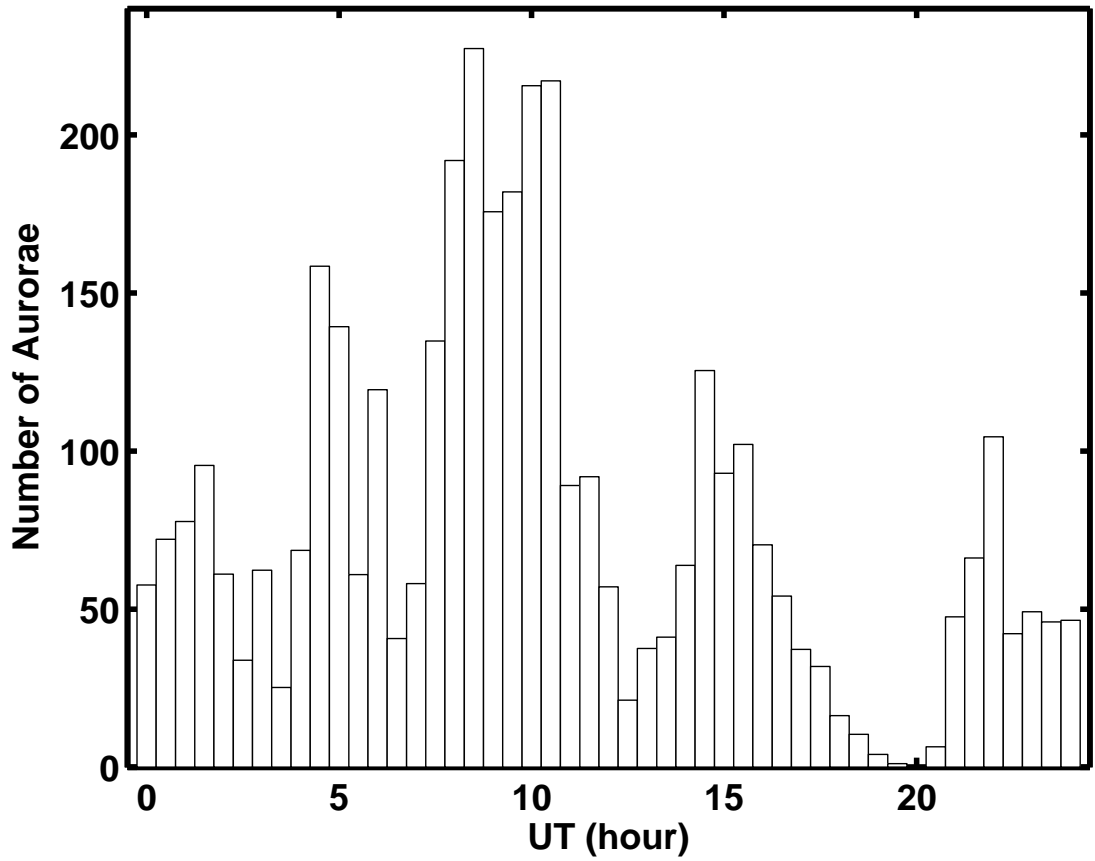} \caption{Time distribution of the
images taken during 2008 that were affected by aurorae. The counts are corrected for the
image-sampling time distribution.\label{fig13}}
\end{figure}

There are about 3800 images polluted by aurorae, or less than 2\% of the total.
Most of the detected aurorae should be comparatively strong ones. The aurorae
with weaker intensities are immersed in the fluctuation of the real observed
data and errors of the fitted model. Therefore, the fraction of aurorae is
a lower limit. However, we believe that any aurorae possibly present in
the remaining 98\% of our images are sufficiently faint as to not affect
photometric accuracy. We do not find obvious dependence of the
aurorae on the transparency variation or cloud coverage, which implies that
the above aurorae are detected reasonably, because the transparency variation
distribution of those polluted images are generally uniform as we check.
Figure \ref{fig13} shows the time distribution of the images affected
by aurorae. A considerable number of them concentrate before or after
7:00 in UT when the Sun is highest. We can also find that there are few or no images
affected by aurorae at about 20:00 nearly when the Sun elevation is
lowest at 19:00.

An optical spectrometer called Nigel \citep{ken06b} was installed in PLATO at Dome A
in 2009 January and this will allow unambiguous identification of aurorae.

\section{CONCLUSION}

In 2008, CSTAR was deployed at Dome A in Antarctica to measure the
night sky background, the weather conditions, and to study variable
stars. About 310,000 images were acquired during the observing period
of over 4 months. We have used these data to quantify the $i$-band sky
brightness, variations in sky transparency, the background flux from
the Sun and the Moon, and to derive some statistics on aurorae.

Due to the high ecliptic latitude of the observed South Celestial
Pole area, the sky background is little affected by the zodiacal
light. The $i$-band sky brightness has a
median value of 20.5 mag arcsec$^{-2}$ when not affected by the Sun and the Moon.
When corrected for the diffuse Galactic background, the sky brightness drops a further 0.06 mag.

Comparing each image to our ``best'', or reference, image observed on
a clear and moonless night, variations in transparency can be
calculated. We find that the weather was stable and there was little cloud during
most of the observing time. Conditions became somewhat worse and the
transparency changed considerably in June and July when the Sun was at
its lowest below the horizon in the year. Among the 83\% images in
which the elevation of the Sun was lower than $-13^\circ$, about
67\% of them were taken under conditions of little or no cloud, using the same criterion for
the cloud cover as used by the Gemini Observatory at Mauna Kea.

On the whole, Dome A has excellent optical sky backgrounds and had low
cloud fractions during 2008. About 2\% of our images were affected by
relatively strong aurorae.

\section*{ACKNOWLEDGMENTS}

This study has been supported by the Chinese National Natural
Science Foundation through grants 10873016, 10803007, 10473012,
10573020, 10633020, 10673012, and 10603006, and by the National Basic
Research Program of China (973 Program), No.~2007CB815403.
This research is also supported by the Chinese PANDA International
Polar Year project and the Polar Research Institute of
China (PRIC). The support of the Australian
Research Council and the Australian Antarctic Division for the PLATO
observatory is gratefully acknowledged.
The authors thank all members of the 2008 and 2009 PRIC
Dome A expeditions for their heroic effort in reaching the site
and for providing invaluable assistance to the expedition astronomers
in setting up and servicing the PLATO observatory and its associated
instrument suite. Iridium communications were provided by the US National
Science Foundation and the United States Antarctic Program. Additional
financial contributions have been made by the institutions involved in
this collaboration.

We thank K.~Mattila for helpful discussions on the diffuse
Galactic light at the South Celestial Pole.

\clearpage

\clearpage

\begin{table}
\begin{center}
\caption{Hours Per Year at Different Sites When the Sun is Below a
Given Elevation. \label{tab1}}
\begin{tabular}{lccccc}
\tableline\tableline
Site & Latitude & Elevation (m) & Hours ($< -18^\circ$) & Hours ($< -13^\circ$) & Hours ($< 0^\circ$) \\
\tableline
Dome A (Antarctica) &$-80^\circ 22^\prime$&  4093 &  1680 & 2606 &  4471 \\
Mauna Kea (Hawaii) &$+19^\circ 50^\prime$& 4194 & 3395 & 3689 &  4394\\
Cerro Pach\'on (Chile) &$-30^\circ 14^\prime$& 2722 & 3330 & 3640 & 4404 \\
La Palma (Spain) &$+28^\circ 46^\prime$& 2332 & 3313 & 3623 & 4381 \\
\tableline
\end{tabular}
\end{center}
\end{table}

\begin{table}
\begin{center}
\caption{Cloud Cover Conditions at Dome A. \label{tab2}}
\begin{tabular}{cccc}
\tableline\tableline
Flux & Excess Extinction (mag) & Fraction & Cloud Cover\\
\tableline
$< 50\%$ & $> 0.75$ & 9\% & Thick \\
 50\%--75\% & 0.31--0.75 & 17\% & Intermediate \\
 75\%--90\% & 0.11--0.31 & 23\% & Thin \\
 $> 90$\% & $< 0.11$ &  51\% & Little or none \\
\tableline
\end{tabular}
\tablecomments{The first column is the observed flux of our standard stars relative to
the reference image. The second column gives the $i$-band magnitude
change (excess extinction) corresponding to the ``flux''. The last column gives the description of the
cloud cover.}
\end{center}
\end{table}

\begin{table}
\begin{center}
\caption{The Comparison of Cloud Cover Between Mauna Kea and Dome A. \label{tab3}}
\begin{tabular}{ccc|c}
\tableline\tableline
\multicolumn{3}{c|}{Mauna Kea (Gemini)} & Dome A\\
\tableline
Cloud Cover & Extinction ($V$) & Fraction & Fraction \\
\tableline
Any other usable & $> 3$ & 10\% & 0 \\
Cloudy & 2--3 & 20\% & 2\% \\
Patchy cloud & 0.3--2 & 20\% & 31\% \\
Photometric & $< 0.3$ & 50\% & 67\%  \\
\tableline
\end{tabular}
\tablecomments{The definition of cloud cover is adopted from the
Gemini Observatory.
For comparison, we use $V - i = 0.07$ in extinction
for the different tranparencies of these two bands as presented in the text.
Note that the term `photometric' as used here is just
one kind of cloud cover category and it is different from the normal term
`photometric night'.}
\end{center}
\end{table}

\end{document}